\documentclass[runningheads]{llncs}
\usepackage{marvosym}
\usepackage{verbatim}
\usepackage{listings}
\usepackage{courier}
\usepackage{graphicx}
\usepackage{amsmath,amssymb,amsfonts}
\usepackage{multirow}
\usepackage{tikz}
\usepackage{array}
\usepackage{stmaryrd}
\usepackage[ruled,vlined,linesnumbered]{algorithm2e}

\usepackage{tikz,xcolor,hyperref}

\definecolor{lime}{HTML}{A6CE39}
\DeclareRobustCommand{\orcidicon}{%
	\begin{tikzpicture}
		\draw[lime, fill=lime] (0,0) 
		circle [radius=0.16] 
		node[white] {{\fontfamily{qag}\selectfont \tiny ID}};
		\draw[white, fill=white] (-0.0625,0.095) 
		circle [radius=0.007];
	\end{tikzpicture}
	\hspace{-2mm}
}

\foreach \x in {A, ..., Z}{%
	\expandafter\xdef\csname orcid\x\endcsname{\noexpand\href{https://orcid.org/\csname orcidauthor\x\endcsname}{\noexpand\orcidicon}}
}

\begin{document}
\title{Efficient Local Search for Nonlinear Real Arithmetic}
\titlerunning{Efficient Local Search for Nonlinear Real Arithmetic}
\author{Zhonghan Wang\orcidA{}$^{1,2}$,
Bohua Zhan$^{\textrm{(\Letter)}}$\orcidB{}$^{1,2}$,
Bohan Li\orcidC{}$^{1,2}$,
Shaowei Cai\orcidD{}$^{1,2}$}
\authorrunning{Z. Wang, B. Zhan, et al.}
\institute{$^1$State Key Laboratory of Computer Science, Institute of Software, \\
    Chinese Academy of Sciences, Beijing, China \\
  \email{\{wangzh,bzhan,libh,caisw\}@ios.ac.cn} \\
$^2$University of Chinese Academy of Sciences, Beijing, China
}
\maketitle              % typeset the header of the contribution
\begin{abstract}

Local search has recently been applied to SMT problems over various arithmetic theories. Among these, nonlinear real arithmetic poses special challenges due to its uncountable solution space and potential need to solve higher-degree polynomials. As a consequence, existing work on local search only considered fragments of the theory. In this work, we analyze the difficulties and propose ways to address them, resulting in an efficient search algorithm that covers the full theory of nonlinear real arithmetic. In particular, we present two algorithmic improvements: incremental computation of variable scores and temporary relaxation of equality constraints. We also discuss choice of candidate moves and a look-ahead mechanism in case when no critical moves are available. The resulting implementation is competitive on satisfiable problem instances against complete methods such as MCSAT in existing SMT solvers.  

\end{abstract}

\section{Introduction}

Satisfiability Modulo Theories (SMT) is the problem of determining the satisfiability of a formula containing both logical operators and functions interpreted in one or more custom theories~\cite{BarrettT18}. Commonly considered theories include equality, arithmetic, bit-vectors, arrays, and strings. After nearly two decades of development, SMT has gained widespread applications in program verification, model checking, planning, and many other areas.

The arithmetic theories can be divided according to the type of numbers involved into integer and real theories, and according to the operations allowed into difference logic, linear, and nonlinear theories. The case of (quantifier-free) nonlinear real arithmetic (NRA) considers satisfiability of equalities and inequalities involving polynomials of degree greater than one, and where the arithmetic variables take on real values. It has applications in the analysis of nonlinear hybrid automata~\cite{CimattiMT12}, generating ranking functions for termination analysis~\cite{HeizmannHLP13,LeikeH15}, constraint answer set programming~\cite{SusmanL16,ShenL18}, and even analysis of biological networks~\cite{AkutsuHT08}. Problem instances from many of these applications are collected in the SMT-LIB benchmarks~\cite{BarFT-SMTLIB}.

Similar to the SAT case, methods for solving SMT problems can be roughly divided into \emph{complete} and \emph{incomplete} methods. Complete methods are usually based on DPLL($T$) or close variants. They are able to both find solutions and prove unsatisfiability. Incomplete methods, such as those based on local search~\cite{HoosS2004}, explore the solution space heuristically, usually by changing the assignment of one variable at a time, in an attempt to find a satisfying solution. Local search methods are not able to prove unsatisfiability, but can have an advantage over complete methods on some satisfiable instances.

Complete methods for the theory of nonlinear real arithmetic make crucial use of cylindrical algebraic decomposition (CAD)~\cite{Caviness2004QuantifierEA}, which permits deciding the satisfiability of a conjunction of polynomial (in)equalities over real numbers. This can then serve as the theory solver in DPLL($T$)~\cite{NieuwenhuisOT06}. An innovation over DPLL($T$) for arithmetic theories is the MCSAT algorithm~\cite{JovanovicM12,MouraJ13}, which constructs models involving both boolean and arithmetic variables at the same time. A nice overview of DPLL($T$) and MCSAT for nonlinear real arithmetic can be found in Kremer's thesis~\cite{Kremer20}.

Exploration of applying local search to solve SMT problems over arithmetic theories began only recently. The work~\cite{CaiLZ22} applied local search to the theory of linear integer arithmetic. It introduced the concept of \emph{critical moves}, a change in one arithmetic variable that satisfies a previously unsatisfied clause. The algorithm iteratively applies critical moves that most improves the score, a weighted count of unsatisfied clauses. The presence of both boolean and arithmetic variables are dealt with by alternately working in the \emph{integer mode} and the \emph{boolean mode}, when assignments to only integer or boolean variables are changed respectively. Switching between modes are performed after the number of non-improving steps reaches a certain threshold.

Compared to linear integer arithmetic, the problem of nonlinear real arithmetic held additional challenges for local search methods. Unlike integer theories, there is an uncountable number of possible assignments to choose from, including infinite number of choices in any finite interval. Unlike the linear case, there is potential need to solve polynomials of degree greater than one, which is both costly in time and may result in variable assignments that are irrational (i.e. algebraic) numbers, causing a slow-down of the ensuing search process. It is also possible that a nonlinear constraint cannot be satisfied by changing the value of one variable alone, resulting in the lack of critical moves, so that other heuristics are needed in such scenarios.

There has already been some work exploring local search for nonlinear real arithmetic. However, largely due to the challenges listed above, none of the existing work covers the entirety of the theory. The work~\cite{abs-2303-06676} considers the multilinear case, where each variable appears with degree at most one in each polynomial constraint. The work~\cite{LiXZ23} considers problems where all equality constraints contain at least one variable that is linear. In both of these works, the problem of variable assignments to irrational values is avoided by either assuming linearity in each variable, or by limiting higher-degree constraints to strict inequalities only.

\subsection{Contributions}

In this paper, we propose several improvements to the local search procedure, aimed at addressing the challenges posed by nonlinear real arithmetic. This results in an efficient local search algorithm for the entire NRA theory.

First, we present efficient data structures for caching and updating variable scores used to determine the next critical move. Existing work on local search for arithmetic theories can be thought of as an extension of the GSAT algorithm~\cite{SelmanLM92} with adaptive weighting. It is well-known that efficient implementations of GSAT involve caching and updating of variable scores~\cite[Section 6.2]{HoosS2004}. For arithmetic theories, this is complicated by the fact that each variable is associated not with one score, but with different scores for changing its assignment to values in different intervals. Hence, current implementations of local search for arithmetic theories recompute the score information for each variable at every iteration, resulting in potential repeated computations. This is especially serious for nonlinear arithmetic, where computations may involve costly root-finding for higher-degree polynomials. We describe data structures maintaining \emph{boundaries} of score changes for each pair of clause and variable appearing in the clause, which only need to be updated on an as-needed basis, and from which the full score information can be quickly recovered.

Second, we address the problem posed by nonlinear equality constraints between variables, which may force assignment to variables that are irrational numbers. Rather than making such assignments directly, we propose to temporarily \emph{relax} such equalities into inequalities, e.g. changing the constraint $p=0$ into $p>-\epsilon \wedge p<\epsilon$, and continue the local search process. If an (approximate) solution is found that satisfies the relaxed version of these constraints, we restore the equalities to their original form, and try to find an exact solution near the approximate solution. For this step, we try two different heuristic methods, respectively based on analyzing the structure of equations and local search itself. While neither method is guaranteed theoretically (or in practice) to find an exact nearby solution every time, the use of relaxation means local search mainly works with rational assignments, significantly improving its efficiency in those problem instances where irrational assignments may otherwise be needed.

Finally, we present alternative ways to deal with situations where no critical move is available to satisfy a certain literal (that is, when the literal is \emph{stuck}). We first pick heuristically a variable appearing in the literal, and a set of candidate moves for that variable. For each candidate move, we look ahead to see whether that move will lead to the literal having critical moves on the next step. Such moves are then preferred over the others.

The above ideas are implemented as a local search algorithm on top of the Z3 prover~\cite{MouraB08}. %\footnote{Binary file of the program for Linux operating system is available at the anonymized repository \url{https://anonymous.4open.science/r/z3_binary-1038/}.}.
The implementation relies on the existing library of polynomials and algebraic numbers in Z3, but is otherwise independent from its implementation of MCSAT for nonlinear real arithmetic. We perform thorough experiments on the SMT-LIB benchmarks, showing the effect of incremental computation of variable scores and relaxation of equalities, and that the resulting local search algorithm is competitive against complete algorithms in existing SMT solvers on the satisfiable instances. Especially, there is a large amount of complementarity between the problems solved by local search and by the complete algorithms, indicating major improvements can be obtained in a portfolio setting.

\vspace{-2mm}
\subsection{Related work}

Our work builds upon existing work applying local search to SMT over arithmetic theories~\cite{CaiLZ22,abs-2303-06676,LiXZ23}. They will be reviewed in more detail in Section~\ref{sec:preliminaries}. Besides arithmetic theories, there have also been earlier work applying local search to the theory of bit-vectors~\cite{GriggioPST11,FrohlichBWH15,NiemetzPB16}. In particular, the work~\cite{FrohlichBWH15} generalized the scoring function to consider operators on bit-vectors, but the moves remain single-bit flips. Later works~\cite{Niemetz2015ImprovingLS,NiemetzPB16} introduced propagation-based move selection and essential inputs to prune the search.

The most commonly used framework for complete algorithms for nonlinear real arithmetic is MCSAT, first proposed by Jovanovic and de Moura~\cite{JovanovicM12,MouraJ13}. It improves upon the use of CAD within the DPLL($T$) framework by assigning to both arithmetic and boolean variables during the search process. Recent innovations in SMT solving for nonlinear arithmetic include variants to the application of CAD~\cite{AbrahamDEK21}, and alternative heuristics for choosing variable ordering~\cite{LiXZZ23}. Besides complete methods based on CAD, alternative methods based on linearization~\cite{CimattiGIRS18}, interval constraint propagation~\cite{KhanhO12,TungKO17}, and subtropical methods~\cite{FontaineOSV17,NalbachA23} are also explored. While these approaches alone do not achieve the same level of overall performance as CAD and MCSAT, they may have advantages in specific classes of problems, making them useful in a portfolio setting.

Also closely-related to our work are various methods for determining whether a given region contains a solution to a set of (in)equality constraints, and their applications to SMT solving. The work of Cimatti et al.~\cite{CimattiGLS22} first uses global optimization methods to find an initial solution, then uses topological methods to determine whether an exact solution exists nearby. The work of Ni et al.~\cite{NiWX23} also uses optimization to find a candidate solution, followed by general methods for solving equations~\cite{LiXZ23b} to isolate an exact solution.

The work of Gao et al.~\cite{GaoAC12IJCAR,GaoAC12LICS} introduced the framework of $\delta$-complete decision procedures, implemented in the \textsf{dReal} tool for solving SMT problems over nonlinear formulas~\cite{GaoKC13}. It can handle polynomials as well as trigonometric and exponential functions. The $\delta$-complete framework allows algorithms to either return $\delta$-$\mathsf{sat}$ or $\mathsf{unsat}$, where the $\delta$-$\mathsf{sat}$ case returns a solution for a $\delta$-weakening of the input formulas. This permits efficient numerical algorithms to be used, as well as showing decidability for a wide range of problems. The concept of relaxation of constraints in our work is similar to $\delta$-weakening, and we also use it to increase efficiency of our algorithm. However, we still aim to return an exact answer, by restoring the constraints to their original form and try to find an exact solution near any approximate solution that is found.

\subsection{Structure of the paper}

We begin by defining the SMT problem over nonlinear real arithmetic, and reviewing existing local search algorithms in Section~\ref{sec:preliminaries}. Section~\ref{sec:incremental-computation} presents incremental computation of variable scores. Section~\ref{sec:relaxation-equalities} presents relaxation and restoring of equality constraints. Section~\ref{sec:implementation} discusses implementation choices, including heuristic move selection and look-ahead mechanism for stuck literals. Section~\ref{sec:evaluation} compares with existing SMT solvers, and performs ablation study on each of the proposed improvements. Finally, we conclude in Section~\ref{sec:conclusion} with a discussion of potential future directions.

\paragraph{Acknowledgements} This work was partially supported by the National Natural Science Foundation of China under Grant Nos. 62032024 and 62002351.

\section{Preliminaries}
\label{sec:preliminaries}

In this section, we formally define SMT problems over nonlinear real arithmetic, followed by a review of existing local search algorithms for arithmetic theories.

The syntax of a general SMT formula over nonlinear real arithmetic is as follows:
\begin{align*}
    p &:= x ~|~ c ~|~ p + p ~|~ p \cdot p &\quad\mbox{(polynomials)} \\
    a &:= b ~|~ p \ge 0 ~|~ p \le 0 ~|~ p = 0 &\quad\mbox{(atoms)} \\
    f &:= a ~|~ \neg f ~|~ f \wedge f ~|~ f \vee f &\quad\mbox{(formulas)}
\end{align*}

Here $x$ is an arithmetic variable, $c$ is a constant rational value, $b$ is a boolean variable. A \emph{literal} is either an atom or its negation. A \emph{clause} is a disjunction of literals. In other words, we consider (in)equalities on polynomials with rational coefficients\footnote{Our methods can be extended to handle polynomials with coefficients that are algebraic numbers. Alternatively, a coefficient $c_i$ that is an algebraic number can be encoded as a variable $x_i$ satisfying some polynomial $p(x_i)=0$ together with interval constraints. Hence we limit the discussion to rational coefficients in this paper.}. In practice, we assume that input problem instances are given in conjunctive normal form (CNF), that is as a collection of clauses to be satisfied. Note that strict inequalities $p\neq 0$, $p<0$ and $p>0$ can be represented as $\neg (p=0)$, $\neg (p\ge 0)$ and $\neg (p\le 0)$, respectively. We allow problem instances to contain boolean and arithmetic variables at the same time. We define \emph{boolean literal} and \emph{arithmetic literal} to mean literal whose atom is a boolean variable and a polynomial inequality, respectively.

A polynomial $p$ is \emph{linear} in some variable $x$ if all terms of the polynomial have degree at most one in $x$. Alternatively, $p$ can be written in the form $p_1\cdot x + p_2$, where $p_1,p_2$ do not contain $x$. A polynomial is \emph{multilinear} if it is linear in each of its variables.

Given a problem instance containing boolean variables $b_i~(1\le i\le m)$ and arithmetic variables $x_j~(1\le j\le n)$, a \emph{complete assignment} is a mapping from each $b_i$ to $\{\top, \bot\}$ and each $x_j$ to $\mathbb{R}$. We will only deal with complete assignments in this paper, and hence sometimes call it \emph{assignment} for short. A formula is satisfied under an assignment if it evaluates to true under the standard interpretation of boolean and arithmetic operators. An assignment is a solution to a problem instance if it satisfies all its clauses. The SMT problem for nonlinear real arithmetic is to determine whether a given problem instance is satisfiable by some assignment.

\emph{Local search} algorithms attempt to determine satisfiability of a problem instance by searching in the space of complete assignments, usually by changing the value of one variable at a time. In the SAT case, each move in local search flips the assignment of one boolean variable. Which flip to make is determined by factors such as the number of clauses that become satisfied/unsatisfied as result of the flip, weight of the clauses, which variables are flipped recently, and so on. For SMT over arithmetic theories, the analogous move changes the value of one arithmetic variable, usually in order to make some clause become satisfied. Such moves, called \emph{critical moves}, are introduced in~\cite{CaiLZ22} for linear integer arithmetic.

For nonlinear real arithmetic, the basic procedure used to determine possible moves is \emph{root isolation} for polynomials. Given a polynomial $p$ in a single variable $x$ (where here we allow the coefficients of $p$ to be algebraic numbers), the procedure computes the roots of the equation $p(x)=0$, together with the sign of the polynomial in each interval separated by the roots. Algebraic numbers in the coefficients of $p$ and in the output of root isolation are represented by their minimal polynomials, together with intervals with rational endpoints that bracket a root of that polynomial.

Given a complete assignment, and an arithmetic literal $l$ involving variable $x$, we can use root isolation to compute the set of values that assignment of $x$ may be moved to in order for $l$ to be satisfied. This is done by substituting in assignments to other variables in $l$, resulting in a polynomial containing only variable $x$, then perform root isolation and check the value of $l$ in each resulting interval. The answer is given in terms of a set of intervals (which may contain $\pm\infty$ as one of the endpoints, and may be either open or closed at each endpoint). We state the definitions precisely as follows.

\begin{definition}
Given a complete assignment, a literal $l$ and a variable $x$, the \emph{feasible set} (resp. \emph{infeasible set}) is the collection of intervals that the value of $x$ can be moved to in order for $l$ to be satisfied (resp. unsatisfied). Likewise, we define the feasible set (resp. infeasible set) of a clause with respect to a variable. This is computed by taking the union (resp. intersection) of the feasible set (resp. infeasible set) for each literal in the clause.
\end{definition}

A critical move is defined to be a change in the assignment of some variable $x$ to a value that satisfies some previously unsatisfied clause. The basic local search algorithm then performs critical moves at each iteration, using various scoring metrics to determine which move is chosen next. Such moves can also be interpreted as jumping between CAD cells as in~\cite{LiXZ23}. An innovation in~\cite{LiXZ23} is that when no critical moves are available for some literal, moves that change multiple variables at once along some straight line are also explored.

Scoring of critical moves is usually based on a weighted count of unsatisfied clauses. \emph{Adaptive weighting schemes} assign a weight to each clause, reflecting its importance during the current search. Existing work on local search for arithmetic theories mostly use a probabilistic version of the PAWS weighting scheme~\cite{ThorntonPBF04}. This scheme is parameterized by a smoothing probability \textit{sp}. Whenever there is no moves available that improves the score, with probability $1-\mathit{sp}$ the weight of each unsatisfied clause is increased by 1, and with probability $\mathit{sp}$ the weight of each satisfied clause with weight greater than 1 is decreased by 1. Then, the \emph{make-break score} of each critical move equals the total weight of clauses that become satisfied by the move, minus the total weight of clauses that become unsatisfied by the move. A move is \emph{improving} if its score is greater than zero.

One key contribution of~\cite{abs-2303-06676} is the introduction of \emph{make-break intervals}. The idea is that instead of considering only the (in)feasible intervals of a variable $x$ with respect to some clause, we combine the (in)feasible interval information of $x$ with respect to all clauses. This results in a partition of the real line into intervals, with each interval associated to the make-break score for moving the value of $x$ into that interval. We illustrate this idea with the following example.

\begin{example}
\label{ex:make-break}
Consider the set of clauses $x^2+y^2\le 1$, $x+y<1$ and $x+z>0$. The current assignment is $x\mapsto 1, y\mapsto 1, z\mapsto 1$, and the current weight of clauses are $1,3,2$, respectively. The make-break score for variable $x$ with respect to each of the clauses are:
\begin{itemize}
    \item $x^2+y^2\le 1$ (unsatisfied): $(-\infty,0)\mapsto 0$, $[0,0]\mapsto 1$, $(0,\infty)\mapsto 0$.
    \item $x+y<1$ (unsatisfied): $(-\infty,0)\mapsto 3$, $[0,\infty)\mapsto 0$.
    \item $x+z>0$ (satisfied): $(-\infty,-1]\mapsto -2$, $(-1,\infty)\mapsto 0$.
\end{itemize}
Combining the above information, we obtain the following make-break intervals and scores for $x$: $(-\infty,-1]\mapsto 1$, $(-1,0)\mapsto 3$, $[0,0]\mapsto 1$, $(0,\infty)\mapsto 0$. A preferred move would be to change the value of $x$ into the interval $(-1,0)$, satisfying the clause $x+y<1$ and leaving the status of the other clauses unchanged, with a make-break score of 3.
\end{example}

If boolean variables are present, the make-break score of flipping each boolean variable is defined in a similar way, as the total weight of clauses that become satisfied by the flip, minus the total weight of clauses that become unsatisfied.

Algorithm~\ref{alg:basic} shows the structure of the basic local search procedure. Begin by initializing the assignments to all variables (line 1). At each iteration, first try to find a move with the largest make-break score. If the score is greater than 0 (the variable necessarily comes from an unsatisfied clause), then perform the corresponding move (line 9). If no move has score greater than 0, it indicates that we reached a local minimum. Update the clause weights according to the PAWS scheme (line 11), and then try to make a move that makes a randomly chosen clause satisfied (line 16). If that is also not possible after several tries, randomly change the assignment of some variable in some unsatisfied clause according to some heuristic (line 19). This continues until all clauses are satisfied (line 4) or the time or step limit is reached (line 6).

\begin{algorithm}[!t]
\caption{Basic local search algorithm}
\label{alg:basic}
\SetKwInOut{Input}{Input}
\SetKwInOut{Output}{Output}
\Input{A set of clauses $F$}
\Output{An assignment of variables that satisfy $F$, or failure}
Initialize assignment to variables\;
\While{$\top$}{
    \If{all clauses satisfied}{
        \Return{success with assignment\;}
    }
    \If{time or step limit reached}{
        \Return{failure\;}
    }
    $\mathit{var,new\_value,score} \leftarrow$ best move according to make-break score\;
    \If{score $>$ 0}{
        Perform move, assigning $\mathit{var}$ to $\mathit{new\_value}$\;
    }
    \Else{
        Update clause weight according to PAWS scheme\;
        \Repeat{3 times}{
            $\mathit{cls} \leftarrow$ random unsatisfied clause\;
            $\mathit{var,new\_value,score} \leftarrow$ critical move making $\mathit{cls}$ satisfied\;
            \If{score $\neq$ $-\infty$}{
                Perform move, assigning $\mathit{var}$ to $\mathit{new\_value}$\;
            }
        }
        \If{no move performed in previous loop}{
            Change assignment of some variable in some unsatisfied clause\;
        }
    }
}
\end{algorithm}

There are possible variations in the choice between boolean and arithmetic variables on line 7. In~\cite{CaiLZ22,abs-2303-06676}, the search is separated into modes where only boolean or arithmetic variables are considered. Alternatively, we can combine the lists of moves and decide between them based purely on make-break scores. We take the latter approach in this paper. There are other aspects of the algorithm that are left unspecified, including how the best move is computed on line 7, and the heuristic choice of moves on line 19. These will be specified in more detail in the next sections.

\section{Incremental computation of variable scores}
\label{sec:incremental-computation}

One key step in Algorithm~\ref{alg:basic} is computing the move with the best make-break score. The computation for boolean variables is standard (and is in any case not the bottleneck here), hence we focus on critical moves for arithmetic variables. The default approach is to loop over all variables in all unsatisfied clauses. For each variable, compute its score with respect to each clause and then combine the results (as demonstrated in Example~\ref{ex:make-break}). However, it is clear that this may result in repeated computations across iterations. For example, the feasible set of some variable with respect to a clause may be recomputed, even if none of the variables in that clause are changed in the previous step. Following the idea of caching and updating scores in GSAT~\cite{SelmanLM92}, we propose data structures for caching and updating score information for arithmetic variables.

We define a \emph{boundary} to be a quadruple $\langle \mathit{val}, \mathit{is\_open}, \mathit{is\_make}, \mathit{cid}\rangle$, where $\mathit{val}$ is a real number, $\mathit{is\_open}$ and $\mathit{is\_make}$ are boolean values, and $\mathit{cid}$ is a clause identifier. The boundary indicates that there is a change in make-break score when moving from less than to greater than $\mathit{val}$ due to clause $\mathit{cid}$. If $\mathit{is\_make}$ is $\top$, the score is increased by the weight of clause $\mathit{cid}$, otherwise it is decreased by the weight. If $\mathit{is\_open}$ is $\top$, the change is not active at $\mathit{val}$, otherwise it is already active at $\mathit{val}$. There is a natural ordering among boundaries, first order by $\mathit{val}$ and then by $\mathit{is\_open}$ (with $\bot<\top$). The make-break score information of each variable with respect to each clause can be characterized by a starting score (indicating the make-break score of large negative values), together with a set of boundaries. The make-break information of each variable with respect to all clauses is formed by summing the starting score and taking the union of the sets of boundaries. We illustrate the computations in the following example.

\begin{example}
\label{ex:boundaries}
    Continuing from Example~\ref{ex:make-break}, the starting score and boundary information of variable $x$ with respect to each clause is as follows (we identify the three clauses as 1,2,3, respectively).
    \begin{itemize}
        \item $x^2+y^2\le 1$: starting score 0, boundary set $\{(0,\bot,\top,1), (0,\top,\bot,1)\}$, indicating no change for large negative values, \emph{make} at boundary $[0,\cdots$, followed by \emph{break} at boundary $(0,\cdots$.
        \item $x+y<1$: starting score 3, boundary set $\{(0,\bot,\bot,2)\}$, indicating \emph{make} at large negative values, and \emph{break} at boundary $[0,\dots$.
        \item $x+z>0$: starting score $-2$, boundary set $\{(-1,\top,\top,3)\}$, indicating \emph{break} at large negative values, and \emph{make} at boundary $(-1,\dots$.
    \end{itemize}
    The combined make-break score information is: starting score 1, with the following (ordered) set of boundaries: $\{(-1,\top,\top,3),(0,\bot,\top,1),(0,\bot,\bot,2),(0,\top,\bot,1)\}$. Make-break score information in terms of intervals can be easily recovered from the above, by traversing the boundaries in order, increasing the score by the weight of the clause when encountering a boundary with $\mathit{is\_make}=\top$, and decreasing the score by the weight otherwise.
\end{example}

During local search, after each move of variable $v$ to a new value, only those variables $v'$ that share a clause with $v$ need to have their make-break score information updated (this is analogous to the concept of dependent variables in the SAT case), and moreover boundary information need to be updated for the shared clauses only. This is summarized in Algorithm~\ref{alg:incremental}. The set $S$ collects the set of variables sharing a clause with $v$. Line 5 recomputes starting score and boundary information. Line 7 recomputes best critical move and score for each updated variable.

\begin{algorithm}[!t]
\caption{Incremental computation of make-break scores}
\label{alg:incremental}
\SetKwInOut{Input}{Input}
\SetKwInOut{Update}{Update}
\Input{Variable $v$ that is modified}
\Update{Make-break score for all variables}
$S \leftarrow \{\}$ \tcp*{set of updated variables}
\For{clause $\mathit{cls}$ that contains $v$}{
    \For{variable $v'$ appearing in $\mathit{cls}$}{
        add $v'$ to $S$\;
        recompute starting score and boundary of $v'$ with respect to $\mathit{cls}$\;
    }
}
\For{variable $v'$ in S}{
    recompute best critical move and score in terms of boundary information\;
}
\end{algorithm}

\begin{example}
\label{ex:update-boundary}
Continuing from Example~\ref{ex:boundaries}, suppose the move $y\mapsto -2$ is made, making the clause $x+y<1$ satisfied. Then the score information for variable $z$ does not need to be updated, as $y$ and $z$ do not share a clause. The score information for variable $x$ need to be updated for the first two clauses. For clause $x^2+y^2\le 1$ there is no longer any boundaries (no assignment of $x$ can make the clause true), and for clause $x+y<1$ the new starting score and boundary set are 0 and $\{(3,\bot,\bot,2)\}$, respectively. So the overall starting score and boundary set are $-2$ and $\{(-1,\top,\top,3),(3,\bot,\bot,2)\}$.
\end{example}

\begin{remark}
Data structures such as arrays, linked lists, or binary trees can be used to maintain set of boundaries. If the total number of boundaries for each variable is small (as is the case for most of the problem instances in SMT-LIB), arrays or linked lists are sufficient. Otherwise the use of binary trees result in better asymptotic performance for the required operations.
\end{remark}

\begin{remark}
A further optimization can be made: it is not necessary to immediately recompute the boundary information for a variable $v'$ that does not appear in any unsatisfied clause, as such variables will never be chosen either on line 7 or line 14 of Algorithm~\ref{alg:basic}. Instead, flags can be used to mark that boundary information for certain clauses need to be updated for $v'$. When at least one of the clauses containing $v'$ becomes unsatisfied, information for those flagged clauses (as well as other clauses that need to be updated for that step) are updated.
\end{remark}

\section{Relaxation of equalities}
\label{sec:relaxation-equalities}

Equality constraints with degree greater than one pose special difficulty for local search, since it may force assignments of variables to irrational (e.g. algebraic) numbers. For example, for the constraint $x^2+y^2+z^2=1$, with most rational assignments to $x$ and $y$, the assignment to $z$ would be forced to be irrational in order for the constraint to be satisfied. While it is possible to represent and compute with algebraic numbers during local search, the time-cost of such computation is significantly increased. Even without considering algebraic numbers, numbers with increasingly large denominators are also problematic for slowing down the search process.

Both~\cite{abs-2303-06676} and~\cite{LiXZ23} avoid algebraic numbers by either limiting themselves to the multilinear case, or considering only equality constraints with at least one linear variable (and solving only for those linear variables). The work~\cite{abs-2303-06676} further incorporated comparison of size of denominators (as well as absolute value) of potential assignments into the scoring heuristic, in order to keep the complexity of assigned values as low as possible.

We propose a novel approach to address the problem of assignments to irrational values and values with large denominators, that allow the algorithm to be applied efficiently to the full set of nonlinear arithmetic problem instances. The approach still relies on comparing complexity of assigned values, hence we first define it below.

\begin{definition}[Complexity of values]
\label{def:complexity}
We define a preorder $\prec_c$ on algebraic numbers as follows. $x\prec_c y$ if $x$ is rational and $y$ is irrational, or if both $x$ and $y$ are rational numbers, and the denominator of $x$ is less than that of $y$. We write $x \sim_c y$ if neither $x\prec_c y$ nor $y\prec_c x$.
\end{definition}

The relaxation mechanism can be described simply as follows: whenever some equality (or inequality) constraints force an assignment of some variable to a comparatively complex value, those constraints are relaxed before continuing the local search process, so that such assignments never actually occur. The implementation is parameterized by two thresholds. The parameter $\epsilon_v$ (for \emph{variable} threshold) specifies the complexity of assigned values (according to Definition~\ref{def:complexity}) beyond which relaxation of constraints should be applied. The parameter $\epsilon_p$ (for \emph{polynomial} threshold) specifies the amount of relaxation of polynomial constraints. Both $\epsilon_v$ and $\epsilon_p$ are chosen to be $10^{-4}$ in the implementation.

It should be noted that the constraints $p\ge 0$ and $p\le 0$ together can also force the assignment of variables to irrational values. These constraints may appear as part of clauses with more than one literal, and hence are not equivalent to $p=0$. This means in general we consider relaxation of non-strict inequalities, although we will still use the slightly imprecise (but more intuitive) description of relaxing equalities throughout the paper. 

The detailed method for determining which constraints to relax is as follows. When computing the best make-break score of a variable $v$, if that score comes from a one-point interval, the set of clause identifiers in the boundaries contributing to that interval are recorded. If the variable $v$ is chosen, and the new value of $v$ is more complex than both $\epsilon_v$ and any other previously assigned value (according to Definition~\ref{def:complexity}), all equalities and non-strict inequalities in the recorded clauses that contribute to the boundary are relaxed. The result of relaxation is as follows.
\begin{itemize}
    \item If the constraint is of the form $p=0$, it is relaxed into the pair of inequalities $p<\epsilon_p$ and $p>-\epsilon_p$.

    \item If the constraint is of the form $p\ge 0$, it is relaxed into $p>-\epsilon_p$. Likewise, if the constraint is of the form $p\le 0$, it is relaxed into $p<\epsilon_p$.
\end{itemize}

Note that strict inequality constraints cannot force a variable to a particular value. After relaxation, the local search process proceeds as before, but with all evaluation of literals and computation of make-break scores according to the relaxed interpretation of literals.

After local search finds a ``solution'' under the relaxation of some constraints, it is only an \emph{approximate} solution. In fact, there is no guarantee that there is an exact solution nearby. We currently try two different ways to find an exact solution near the approximate solution.

The first method performs a heuristic analysis of the structure of the relaxed constraints, in an attempt to find an order of solving for the variables that is likely to produce variable assignments that satisfies all of the equations. Essentially, we are trying to find an exact solution to a set of equality constraints, nearby an existing approximate solution. The analysis performs the following steps:
\begin{enumerate}
\item If any variable is currently assigned to zero, this is substituted into the constraints. We found this very helpful in practice for eliminating many terms in the constraints.
\item Eliminate any variable $x$ in an equation of the form $p\cdot x+q=0$, where the valuation of $p$ under the current assignment is not close to zero. This is more general than eliminating variables during preprocessing, which requires $p$ to be constant (see Section~\ref{sec:implementation-details}).
\item Finally, we iteratively look for a variable that appear only in one of the equations. We associate this variable to the corresponding equation, and then remove the equation from consideration in the ensuing iterations.
\end{enumerate}
If no equations remain at the end of Step 3, we attempt to find an exact solution to the equality constraints by solving for the variables in reverse order of the above process. We begin by considering the association between variables and equations in Step 3 in reverse order, solving for each variable using the associated equation. Then we obtain the values of solved variables in Step 2 in reverse order. The final exact solution to these equations is checked again for satisfaction of all clauses (various numerical inaccuracies may prevent it from being so). If there are unsolved equations remaining after Step 3 of the above analysis, or if the resulting solution fails to satisfy all clauses, we move to the second approach. 

The second approach uses a simplified version of local search itself to try to move the approximate solution to an exact solution. First, restore the relaxed constraints to their original forms, and then proceed with local search on the arithmetic variables only, until either an exact solution is found, or no improvement can be made. This is a more limited form of the general local search, as we do not attempt changes to boolean variables, or try random moves in case no improvements can be made. Hence it is likely to terminate quickly with either an exact solution or report of failure.

The above description is summarized as a modification of the overall algorithm, shown in Algorithm~\ref{alg:relax}. The main change to the search process is to relax constraints whenever it forces an assignment to values that are more complex than the variable threshold (line 18). If all clauses are satisfied (indicating an approximate solution is found), we first try finding an exact solution nearby by analyzing the structure of relaxed constraints (line 4). If this fails, we try the limited form of local search described above (line 8-9). If this also fails, the search is restarted with a fresh assignment (line 13).

In practice, we find the two heuristic approaches for finding exact solutions to be useful in different scenarios. The first approach deals nicely with cases where the structure of equality constraints involves solving systems of linear equations, but otherwise poses no difficult choices. The second approach can better handle those cases where there may be choices in which variables to modify, but only some of them is correct in order to satisfy the other inequality constraints. It may also be possible to apply more advanced methods for solving equations or determining existence of solutions in~\cite{LiXZ23b,CimattiGLS22}. However, the methods we implement easily extends to the non-zero-dimensional case, and returns exact solutions that can be verified independently. We leave the incorporation of more advanced methods for solving equations to future work.

\begin{algorithm}[h!]
\caption{Relaxation of equalities}
\label{alg:relax}
\SetKwInOut{Input}{Input}
\SetKwInOut{Output}{Output}
\Input{A set of clauses $F$}
\Output{An assignment of variables that satisfy $F$, or failure}
Initialize assignment to variables\; 
\While{$\top$}{
    \If{all clauses satisfied}{
        $\mathit{success} \leftarrow$ find exact solution by analyzing structure of equations\;
        \If{success}{
            \Return{success with assignment};
        }
        \Else{
            Restore relaxed constraints to original form\;
            $\mathit{success} \leftarrow$ find exact solution by limited local search\;
            \If{success}{
                \Return{success with assignment};
            }
            \Else{
                Perform major restart\;
            }
        }
    }
    \If{time or step limit reached}{
        \Return{failure}\;
    }
    \If{no improvement for $T_1$ steps}{
        Perform minor restart\;
    }
    Proceed as in line 7-19 of Algorithm~\ref{alg:basic}, except constraints may be relaxed\;
}
\end{algorithm}

\section{Implementation}
\label{sec:implementation}

In this section, we describe the implementation in more detail. First, we explain our choice of heuristic move selection when encountering a literal without critical moves. Then, we describe some further details on preprocessing, restart mechanism, and other efficiency improvements.

\subsection{Heuristic moves selection and look-ahead}

One major difficulty for local search in nonlinear arithmetic is that it is not always possible to find single-variable moves to satisfy a particular constraint. For example, given constraint $x^2+y^2<1$, and the current assignment $x\mapsto 2, y\mapsto 3$, it is not possible to satisfy the constraint by moving only one of $x$ and $y$. During local search, this is reflected by the situation that no critical move is available for a clause or literal.

Solving this problem in general would likely require complex algorithms such as CAD or polynomial optimization. Indeed, one category in the SMT-LIB benchmarks, \textsf{Sturm-MBO}, coming from analysis of biological networks~\cite{AkutsuHT08}, consists exclusively of problems that require a very complex polynomial to evaluate to zero, subject to positivity constraints of the variables. When the problem has many variables (is high-dimensional), any approach based on heuristically searching for assignments would have difficulty finding the exact combination of assigned values required to satisfy the constraint.

One approach is given in~\cite{LiXZ23}, which involves searching in directions other than those parallel to the coordinate axes to look for solutions. The use of gradient information, as well as scoring based on values of polynomials, increase the chance of finding a solution.

In this paper, we propose another approach that still involves moving only one variable at a time. We say a literal is \emph{stuck} if it is currently unsatisfied and has no critical moves to make it satisfied. Given a literal $l$ that is stuck, we first choose a variable $x$ in $l$ whose coefficient is nonzero (according to the current assignment of the other variables), then heuristically pick a set of candidate values to move the assignment of $x$ to. For each candidate value, we compute whether $l$ is still stuck after making that move. We then prefer those moves that result in $l$ no longer being stuck.

Given the current assignment $x_0$ and the feasible set $I$ of variable $x$ (see Section~\ref{sec:implementation-details}), the heuristic move selection include the following:
\begin{enumerate}
    \item rational numbers and integers close to the boundary inside each interval of $I$. The rational numbers are chosen to be within $10^{-4}$ of the boundary.
    \item the next integer smaller or larger than $x_0$.
    \item three numbers chosen uniformly in the interval $[\frac{x_0}{2},x_0)$, and three numbers chosen uniformly in the interval $(x_0,2x_0]$.
\end{enumerate}

The first class reflects what we know about the constraints on $x$. The second class attempts basic random walk, and prefers (simple) integer values. The third class is the most general, allowing search over large/small values as well as fractions. 

The above ideas are summarized in Algorithm~\ref{alg:look-ahead}. The heuristic choice of candidate values are collected into set $S$ (line 2). Then each value in $S$ is tested in turn. If $l$ has critical move after assigning to any value, that value is returned (line 5). Otherwise a randomly chosen value from $S$ is returned (line 7).

\begin{algorithm}
\caption{Heuristic choice of candidate values and look-ahead for critical moves}
\label{alg:look-ahead}
\SetKwInOut{Input}{Input}
\SetKwInOut{Output}{Output}
\Input{Literal $l$ without critical moves}
\Output{Candidate variable $x$ and new value $x_1$}
$x \leftarrow$ variable in polynomial of $l$ with nonzero coefficient\;
$S \leftarrow $ heuristic move selection for variable $x$\;
\For{value $x_1$ in $S$}{
    \If{$l$ has critical move after assigning $x$ to $x_1$}{
        \Return{$x$, $x_1$}
    }
}
$x_1 \leftarrow$ randomly chosen value in $S$\;
\Return{$x$, $x_1$}
\end{algorithm}

\subsection{Implementation details}
\label{sec:implementation-details}

The algorithm is implemented on top of the Z3 prover, and makes use of its library for polynomials and algebraic numbers, as well as data structures for clauses and literals, but otherwise separate from its implementation of the MCSAT algorithm.

\textbf{Preprocessing.} The following preprocessing steps are used. Eliminate clauses with a single boolean variable and propagate assignments. Combine constraints $p\ge 0$ and $p\le 0$ into equality $p=0$ (when they appear as clauses on their own; literals $p\ge 0$ and $p\le 0$ that are parts of larger clauses cannot be combined). Eliminate variable $x$ in an equation of the form $c\cdot x+q=0$, where $c$ is a constant and $q$ is a polynomial with degree at most 1 and containing at most 2 variables. The conditions on $q$ are designed so that preprocessing does not significantly increase the complexity of the remaining clauses.

\textbf{Restart mechanism.} We use a two-level restart mechanism with two parameters $T_1$ and $T_2$ (both chosen to be 100 in our implementation). Perform a \emph{minor restart} after $T_1$ moves without improvements, which randomly changes one of the variables in one of the unsatisfied clauses. After $T_2$ such minor restarts, a \emph{major restart} is performed that resets the value of all variables.

\textbf{Shortcut for linear equations.} Root-isolation is done by calling the existing implementation in Z3, except when the variable to be solved is linear in the polynomial, in which case a direct (and more efficient) solution method is used.

\textbf{Infeasible sets of variables.} For each clause involving a single variable, derive infeasible set for that variable implied by the clause. Experience shows that excluding assignments from the infeasible set during local search is beneficial for some problem instances but not others. Hence, we exclude such assignments on alternate turns of minor restarts.

\textbf{Parameter settings.} Values of tunable parameters used in the implementation are summarized in Table~\ref{tab:parameters}.

\begin{table}[]
    \centering
    \begin{tabular}{c|c|c}
      Symbol & Explanation & ~Value~ \\ \hline
      $\mathit{sp}$ & Probability $\mathit{sp}$ for PAWS scheme & 0.006 \\
      $T_1$ & Number of non-improving steps before minor restart & 100 \\
      $T_2$ & Number of minor restarts before major restart & 100 \\
      $\epsilon_v$ & Threshold for relaxing equality & $10^{-4}$ \\
      $\epsilon_p$ & Amount of relaxation & $10^{-4}$ \\ \hline
    \end{tabular}
    \vspace{2mm}
    \caption{Tunable parameters of the algorithm}
    \label{tab:parameters}
\end{table}

\section{Evaluation}
\label{sec:evaluation}

In this section, we compare the implementation with those of complete procedures in existing SMT solvers Z3~\cite{MouraB08}, cvc5~\cite{BarbosaBBKLMMMN22} and Yices~\cite{Dutertre14}, as well as previous work on local search for (fragments of) nonlinear arithmetic. We also perform an ablation study on the two improvements described in Section~\ref{sec:incremental-computation} and~\ref{sec:relaxation-equalities}.

The benchmark used in the evaluation comes from SMT-LIB's QF\_NRA theory. The benchmark consists mostly of industrial problems from various applications of constraint solving in nonlinear real arithmetic, including analysis of nonlinear hybrid automata (\textsf{hycomp})~\cite{CimattiMT12}, and generating ranking functions for termination analysis~(\textsf{LassoRanker})~\cite{HeizmannHLP13,LeikeH15}. The \textsf{kissing} benchmark contains encoding of the \emph{kissing problem}, which asks how many unit spheres in a given dimension can be placed tangent to a single unit sphere without intersecting each other. Each benchmark is labeled with either \textsf{sat}, \textsf{unsat} or \textsf{unknown}, according to whether it is known to be satisfiable/unsatisfiable at time of submission. Many of the \textsf{unknown} problems are in fact shown to be unsatisfiable by complete algorithms implemented in solvers such as Z3 and cvc5. For the experiments, we choose all problems from the benchmarks that are labeled \textsf{sat} or \textsf{unknown}, but excluding those \textsf{unknown} instances that are found to be unsatisfiable by either Z3 or cvc5. We also note that the SMT-LIB benchmarks contain problems with a wide range of difficulties, but without specified difficulty ratings. For example, many problem instances in the \textsf{metitarski} category, from the MetiTarski tool for proving theorems involving special functions~\cite{AkbarpourP10}, are quite small and do not pose much of a challenge for either local search or complete algorithms.

This yields a total of 6216 instances. The experiments are run on a cluster of machines with Intel Xeon Platinum 8153 processor at 2.00GHz. Each experiment is run with a time limit of 20 minutes (as in the SMT competition) and memory limit of 30GB.

\subsection{Overall result}

Results of our implementation are compared against that of other SMT solvers in Table~\ref{tab:comparison-solvers}. One major advantage of our algorithm is in the \textsf{Sturm-MBO} category, which involves a single complicated polynomial that tripped up other solvers. However, we also showed good result across other categories, and solved most instances overall.

\begin{figure}
    \centering
    \includegraphics[width=0.45\textwidth]{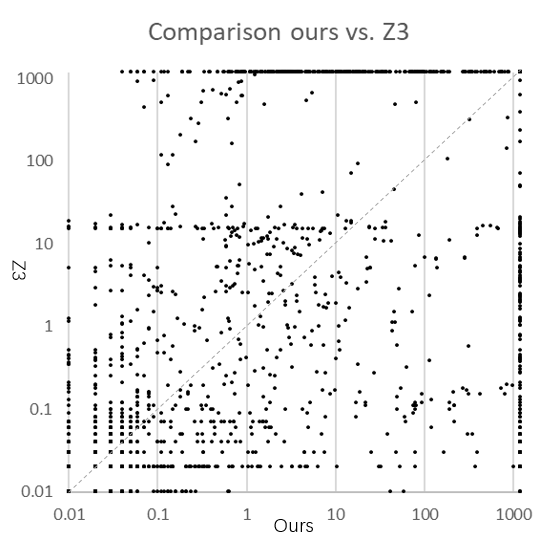}\qquad
    \includegraphics[width=0.45\textwidth]{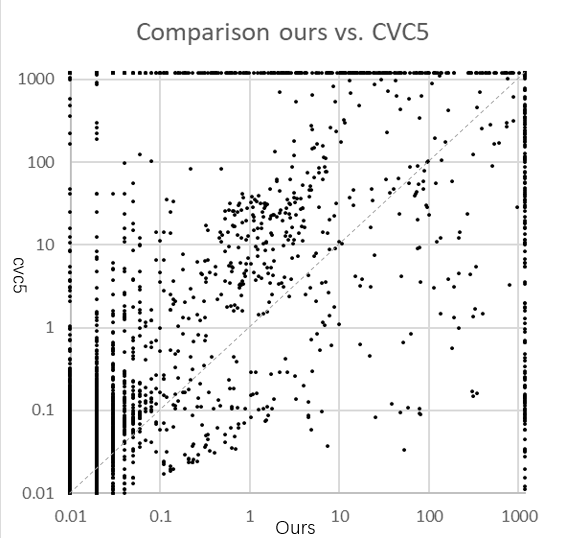}
    \caption{Scatter plots of running time vs. Z3 and cvc5.}
    \label{fig:scatter_plots}
\end{figure}

There is significant amount of complementarity between our algorithm and both Z3 and cvc5. As shown in Table~\ref{tab:comparison-solvers}, there are 148 instances across seven different categories that are solved by local search, but none of Z3, cvc5, and Yices. Moreover, there are 291 instances solved by local search but not Z3, and 378 instances solved by local search but not cvc5. Scatter plots comparing solution times against Z3 and cvc5 are shown in Figure~\ref{fig:scatter_plots}, showing there is significant complementarity in solving times as well.

We also note there is a large number of relatively simple problem instances among the SMT-LIB benchmarks for QF\_NRA. To put this into quantitative form, we counted the number of instances that are solved within 1 second by all of Z3, cvc5, and our solver. There are 4765 such instances, leaving only 1451 instances that can be considered ``challenging'' to the solvers. From this view, the overall improvement 
in the number of solved instances, number of unique solved, and amount of complementarity are quite significant.

\begin{table}[!t]
\small
\centering
\begin{tabular}{c | >{\centering}m{1cm} | >{\centering}m{1cm} | >{\centering}m{1cm} | >{\centering}m{1cm} | c | c}
Category & \#inst & Z3 & cvc5 & Yices & ~Ours~ & ~Unique~ \\ \hline
20161105-Sturm-MBO & 120 & 0 & 0 & 0 & \textbf{88} & 88 \\
20161105-Sturm-MGC & 2 & \textbf{2} & 0 & 0 & 0 & 0 \\
20170501-Heizmann & 60 & 3 & 1 & 0 & \textbf{8} & 6 \\
20180501-Economics-Mulligan & 93 & \textbf{93} & 89 & 91 & 90 & 0 \\
2019-ezsmt & 61 & \textbf{54} & 51 & 52 & 19 & 0 \\
20200911-Pine & 237 & \textbf{235} & 201 & \textbf{235} & 224 & 0 \\
20211101-Geogebra & 112 & \textbf{109} & 91 & 99 & 101 & 0 \\
20220314-Uncu & 74 & 73 & 66 & \textbf{74} & 70 & 0 \\
LassoRanker & 351 & 155 & \textbf{304} & 122 & 272 & 13\\
UltimateAtomizer & 48 & \textbf{41} & 34 & 39 & 27 & 2 \\
hycomp & 492 & \textbf{311} & 216 & 227 & 304 & 11 \\
kissing & 42 & \textbf{33} & 17 & 10 & \textbf{33} & 1 \\
meti-tarski & 4391 & \textbf{4391} & 4345 & 4369 & 4351 & 0 \\
zankl & 133 & 70 & 61 & 58 & \textbf{100} & 27 \\ \hline
Total & 6216 & 5570 & 5476 & 5376 & \textbf{5687} & 148 
\end{tabular}
\vspace{2mm}
\caption{Comparison with other SMT solvers. Column \#inst is the number of instances in the category. Column Z3, cvc5, Yices, and Ours shows number of solved instances by three existing SMT solvers and our implementation. Column Unique is the number of instances solved by our implementation but none of the other three SMT solvers.}
\label{tab:comparison-solvers}
\end{table}

\subsection{Comparison with other work on local search}

We further compare our results against existing work on local search for fragments of nonlinear real arithmetic. Of the 979 instances that are multilinear considered by~\cite{abs-2303-06676}, our implementation can solve 826 instances, compared to 891 solved instances there. The slightly weaker result is likely due to the more efficient implementation that is possible when only rational numbers need to be considered, and the parameter tuning that is specific to multilinear problems. Of the 2736 instances from SMT-LIB considered by~\cite{LiXZ23}, our implementation can solve 2589 instances, compared to 2246 solved instances there. In fact, we solve not only more instances than the local search algorithm given in~\cite{LiXZ23}, but also all other SMT solvers used in the comparison. We note that the result given in~\cite{LiXZ23} uses different underlying software (including Maple) and runs on different machines, so this gives only a rough comparison.

\subsection{Effect of incremental computation of variable scores}

To show the effect of speedup resulting from incremental computation of variable scores in Section~\ref{sec:incremental-computation}, we compare three versions of the implementation: with incremental computation (Incremental), without incremental computation (Naive), and without incremental computation, but limiting the number of unsatisfied clauses considered at each turn to 45 (Limit-45). The results are shown in Table~\ref{tab:compare-incremental}.

We see that while the difference in total number of problem instance solved is not large, a noticeable effect is still present in the \textsf{LassoRanker} category, whose instances usually require a long time to solve. A closer look at the running time shows that it usually takes 2-10 times longer to solve a particular instance using either (Naive) or (Limit-45) compared to (Incremental), with the exact ration depending strongly on the specific instance. For a time limit of 20 minutes the resulting number of solved problems is not large, but we expect a larger difference with shorter time limits, and especially when local search is incorporated into other methods such as DPLL~\cite{CaiZ21,CaiZFB22}.

\begin{table}[!t]
\small
\centering
\begin{tabular}{c | c | c | c | c }
Category & ~\#inst~ & ~Incremental~ & ~Naive~ & ~Limit-45~ \\ \hline
20161105-Sturm-MBO & 120 & 88 & 85 & 85 \\
20161105-Sturm-MGC & 2 & 0 & 0 & 0 \\
20170501-Heizmann & 60 & 8 & 5 & 5 \\
20180501-Economics-Mulligan & 93 & 90 & 89 & 89 \\
2019-ezsmt & 61 & 19 & 19 & 15 \\
20200911-Pine & 237 & 224 & 222 & 222 \\
20211101-Geogebra & 112 & 101 & 101 & 101 \\
20220314-Uncu & 74 & 70 & 70 & 70 \\
LassoRanker & 351 & 272 & 264 & 269 \\
UltimateAtomizer & 48 & 27 & 26 & 26 \\
hycomp & 492 & 304 & 298 & 298 \\
kissing & 42 & 33 & 32 & 33 \\
meti-tarski & 4391 & 4351 & 4352 & 4352 \\
zankl & 133 & 100 & 100 & 100 \\ \hline
Total & 6216 & 5687 & 5663 & 5665
\end{tabular}
\vspace{2mm}
\caption{Comparison showing effect of incremental computation}
\label{tab:compare-incremental}
\end{table}

\subsection{Effect of relaxation of equalities}

We demonstrate the effect of relaxation of constraints by comparing three possible implementations: with relaxation of constraints (Relaxation), without relaxation of constraints, but preferring variable assignments that are less complex than $\epsilon_v$ (Threshold), and without relaxation of constraints, with choosing variable assignments without considering complexity order (NoOrder). The results are shown in Table~\ref{tab:compare-relaxation}.

\begin{table}[!t]
\small
\centering
\begin{tabular}{c | c | c | c | c }
Category & ~\#inst~ & ~Relaxation~ & ~Threshold~ & ~NoOrder~ \\ \hline
20161105-Sturm-MBO & 120 & 88 & 100 & 99 \\
20161105-Sturm-MGC & 2 & 0 & 0 & 0 \\
20170501-Heizmann & 60 & 8 & 9 & 3 \\
20180501-Economics-Mulligan & 93 & 90 & 89 & 86 \\
2019-ezsmt & 61 & 19 & 19 & 19 \\
20200911-Pine & 237 & 224 & 223 & 222 \\
20211101-Geogebra & 112 & 101 & 98 & 92 \\
20220314-Uncu & 74 & 70 & 70 & 70 \\
LassoRanker & 351 & 272 & 277 & 278 \\
UltimateAtomizer & 48 & 27 & 26 & 20 \\
hycomp & 492 & 304 & 211 & 164 \\
kissing & 42 & 33 & 31 & 27 \\
meti-tarski & 4391 & 4351 & 4353 & 4360 \\
zankl & 133 & 100 & 100 & 100 \\ \hline
Total & 6216 & 5687 & 5606 & 5540
\end{tabular}
\vspace{2mm}
\caption{Comparison showing effect of temporary relaxation of constraints}
\label{tab:compare-relaxation}
\end{table}

The results indicate that while taking complexity of assigned values into consideration does have an effect in keeping the search efficient for most categories of problem instances, it is not sufficient for the \textsf{hycomp} category, which involves a large number of nonlinear equalities. In that category using relaxation of constraints have a significant effect, while also performing well in other categories.

\subsection{Other techniques}

We also tried other techniques commonly used in works on local search, including tabu search~\cite{DBLP:books/daglib/0093574}, switching between phases for adjusting boolean and arithmetic variables (as applied in~\cite{CaiLZ22}), and incorporating random walk (such as variants of WalkSAT~\cite{HoosS00}). Unlike their applications in earlier work, the use of such methods did not result in noticeable improvements on the current benchmark. However, it remains to investigate whether they will be useful on other types of problems, or in combination with other improvements to the algorithm.

\section{Conclusion}
\label{sec:conclusion}

In this paper, we presented improvements to the local search algorithm for solving SMT problems in nonlinear real arithmetic. Building upon the basic structure of local search, we presented incremental computation of variable scores and temporary relaxation of constraints. We also described heuristic move selection with look-ahead for dealing with literals without critical moves, and implementation details to improve efficiency. The resulting implementation is competitive against complete algorithms based on DPLL($T$) and MCSAT on satisfiable problem instances, as implemented in other SMT solvers. It is the first local search algorithm designed for the entirety of nonlinear real arithmetic, covering a wider range of problems than existing work~\cite{abs-2303-06676,LiXZ23}.

While the methods proposed in this paper made progress in addressing challenges of local search for nonlinear real arithmetic, there are remaining problems that represent interesting directions of future work.
\begin{itemize}
    \item Look-ahead for critical moves presents another way to improve upon random search in cases when no critical move is available. On the other hand, methods based on CAD or polynomial optimization would give a more complete way to determine assignments that satisfy a certain literal. A major challenge is how to incorporate such algorithms into local search in an efficient way.

    \item In the current work, after an approximate solution is found that satisfies the relaxed version of equalities, we use various heuristic methods to attempt to find an exact solution nearby. Designing or incorporating more general algorithms for finding exact solutions near approximate solutions (or determining that none exist) is an interesting problem that we leave to future work.

    \item Finally, local search can be incorporated into complete methods such as DPLL, improving its performance even on unsatisfiable instances, as shown by the works~\cite{CaiZ21,CaiZFB22}. It is interesting to investigate this possibility for SMT problems over nonlinear real arithmetic. The improvements in efficiency in our work would be very helpful for such combination, as in those cases local search is only given very short running times. 
\end{itemize}

\bibliographystyle{splncs04}
\bibliography{main}

\end{document}